\documentclass[aps,epsf,twocolumn,showpacs]{revtex4}
\usepackage{amsmath}
\usepackage{epsfig}

\begin{document}

\title{Critical Binder cumulant and universality: \\ Fortuin-Kasteleyn clusters and order-parameter fluctuations}

\author{Anastasios Malakis$^1$}

\author{Nikolaos G. Fytas$^2$}

\author{G\"{u}l G\"{u}lpinar$^3$}

\affiliation{$^1$ Department of Physics, Section of Solid State
Physics, University of Athens, Panepistimiopolis, GR 15784
Zografos, Athens, Greece}

\affiliation{$^2$ Applied Mathematics Research Centre, Coventry
University, Coventry, CV1 5FB, United Kingdom}

\affiliation{$^3$ Department of Physics, Dokuz Eyl\"{u}l
University, Buca 35160, Izmir, Turkey}

\date{\today}

\begin{abstract}
We investigate the dependence of the critical Binder cumulant of
the magnetization and the largest Fortuin-Kasteleyn cluster on the
boundary conditions and aspect ratio of the underlying square
Ising lattices. By means of the Swendsen-Wang algorithm, we
generate numerical data for large system sizes and we perform a
detailed finite-size scaling analysis for several values of the
aspect ratio $r$, for both periodic and free boundary conditions.
We estimate the universal probability density functions of the
largest Fortuin-Kasteleyn cluster and we compare it to those of
the magnetization at criticality. It is shown that these
probability density functions follow similar scaling laws, and it
is found that the values of the critical Binder cumulant of the
largest Fortuin-Kasteleyn cluster are upper bounds to the values
of the respective order-parameter's cumulant, with a splitting
behavior for large values of the aspect ratio. We also investigate
the dependence of the amplitudes of the magnetization and the
largest Fortuin-Kasteleyn cluster on the aspect ratio and boundary
conditions. We find that the associated exponents, describing the
aspect ratio dependencies, are different for the magnetization and
the largest Fortuin-Kasteleyn cluster, but in each case are
independent of boundary conditions.
\end{abstract}

 \maketitle

\section{Introduction}
\label{sec:1}

According to the universality
hypothesis~\cite{Fis66,Griff70,kad71}, all critical systems with
the same dimensionality, the same symmetry of the order parameter,
and the same range of interactions are expected to share the same
set of critical exponents. For the two-dimensional (2d) Ising
model (square and some other lattices) all critical exponents are
known exactly~\cite{Ons44,Kauf49,Yang52,Baxt82}. These exponents
are expected to be obeyed by the Ising model on all 2d lattices
and also by all other models, which according to the hypothesis
are expected to belong in the same universality class.
Furthermore, there is strong evidence that, in addition to
critical exponents, certain critical-point ratios are
universal~\cite{Bind81,Priv84,Blot85} and of particular interest
is the value of the critical Binder cumulant of the order
parameter, discussed also in the present work.

The fourth-order cumulant of some thermodynamic parameter $Q$ of a
finite lattice system, known as the Binder cumulant, is defined
as~\cite{Bind81}
\begin{equation}
\label{eq:1}
 U_{\rm Q}(T,L)=1-\frac{\langle Q^{4} \rangle_L}{3\langle
 Q^2\rangle_{L}^{2}},
\end{equation}
with $L$ the linear lattice size. The critical value of the Binder
cumulant of the order parameter of an Ising system is then
\begin{equation}
\label{eq:2} U^{\ast}_{\rm M}=\lim_{L\rightarrow \infty}{U_{\rm
M}(T=T_c,L)},
\end{equation}
with $M$ the magnetization
\begin{equation}
\label{eq:3} M=(1/N)\sum_{i=1}^{N}\sigma_{i},
\end{equation}
$\sigma_{i}$ the spin variable, and $N$ the number of lattice
sites. This parameter is a measure of the deviation of the
universal probability density function from a Gaussian function.
It is well known that the characteristic behavior of $U_{\rm
M}$~\cite{Bind81,BH1988} near criticality provides a traditional
route to obtain transition temperatures (from the intersection of
the cumulants of systems with different sizes) and may also be
used to extract the critical exponent $\nu$ of the correlation
length~\cite{Bind81,Priv84,Blot85}. Its critical value,
$U^{\ast}_{\rm M}$, was originally believed to fully characterize
a given universality class. As discussed by various authors, the
same value seems to be shared by several 2d models, such as the XY
models with an easy axis, the nearest-neighbor spin-1 Ising model,
and the isotropic nearest-neighbor Ising-like models, including
also the nearest-neighbor ``border $\phi^{4}$ model'' with
softened spins~\cite{HSL2005,RPB2005, STTD2002,KB1993, NB1988,
JKV1994, HJMS1997}. It appears also to be independent of the
lattice details, such as the lattice structure~\cite{Selke06,
Selke07}. However, this ``universality'' applies only in a limited
sense. The value of $U^{\ast}_{\rm M}$ does depend on the boundary
conditions~\cite{Bind81,BH1988}, the shape of the
system~\cite{KB1993, JKV1994,SS2005,SS2009,Hilfer2003,DC04,D08},
as well as on the symmetry of the interactions~\cite{Selke06}.

An accurate estimation of critical point ratios for the
ferromagnetic Ising model on the square and triangular lattices
has been provided via the transfer-matrix technique~\cite{KB1993}.
In this paper, Kamieniarz and Bl\"{o}te estimated $U^{\ast}_{\rm
M}$ as a function of the aspect ratio $r$ (see discussion below
for the definition of $r$), reporting, in particular for the
square Ising model with periodic boundary conditions and $r=1$,
the value $U^{\ast}_{\rm M}= 0.61069\cdots$. The influence of the
anisotropic interactions on the critical Binder cumulant was
studied, analytically, by Dohm and Chen~\cite{DC04,D08} and,
numerically, by Selke and Shchur~\cite{SS2005,SS2009}, indicating
that $U^{\ast}_{\rm M}$ depends continuously on the anisotropy, in
the case of periodic boundary conditions and $r=1$. Furthermore,
Kastening~\cite{Kastening} obtained a renormalization-group
quantitative description of the anisotropy dependence of
$U^{\ast}_{\rm M}$.

In the present paper, we investigate certain aspects of the
critical Binder cumulant and, in particular, its dependence on the
boundary conditions and the relevant aspect ratio of the lattice.
We concentrate our interest on numerical observations illustrating
parallel behavior to that of the critical Binder cumulant of the
largest Fortuin-Kasteleyn cluster (LFKC). The rest of the paper is
organized as follows: In the next section, we define the model and
outline the numerical details. Then, in Sec.~\ref{sec:3}, we
present and discuss our numerical findings. Finally, we summarize
our conclusions in Sec.~\ref{sec:4}.

\section{Model and Simulation Details}
\label{sec:2}

Let the size of the LFKC be denoted by $\mathcal{S}_{\rm LFKC}$
and let us define, in analogy to the magnetization,
$l_{\infty}=\mathcal{S}_{\rm LFKC}/N$. Then, the relevant critical
Binder cumulant may be denoted as $U^{\ast}_{\rm l_\infty}$. To be
concrete, we consider the square Ising model in zero field, with
the standard Hamiltonian
\begin{equation}
\label{eq:1}
 \mathcal{H}=-J\sum_{\langle ij \rangle}\sigma_{i}\sigma_{j},
\end{equation}
where the spin variables $\sigma_{i}$ take the Ising values $\pm
1$ and $\langle ij \rangle$ denotes summation over all
nearest-neighbor pairs of sites. For the needs of our study, we
construct ferromagnetic nearest-neighbor square Ising systems with
$L$ rows and $L_1=rL$ columns, corresponding to $N=L\times
L_1=L\times rL\equiv (L^{\ast}){^2}$ sites. Furthermore, we
consider several values of the aspect ratio
$r=\{1,4,9,16,25,36,50,64,100\}$, and investigate both periodic
(PBC) and free boundary conditions (FBC). As our numerical
vehicle, we implement the Swendsen-Wang
algorithm~\cite{Swendsen,Newman99,LandBind00,Ma10}, and we
identify clusters by the Hoshen-Kopelman
procedure~\cite{LandBind00,HoKo}.

The comprehensive Monte Carlo study of De Meo \emph{et
al.}~\cite{Meo} presented a review of the connections of Fortuin
and Kasteleyn's work~\cite{FoKa} to the Swendsen-Wang algorithm
and a review of the relevant literature. In this study, the
authors investigated the scaling properties of the cluster size
distribution and provided a numerical verification of the
theoretical results given by Hu~\cite{Hu84}. In particular, they
showed that the relevant bond-correlated percolation model has the
Ising critical temperature and critical exponents. Thus, it is
generally assumed, that the LFKC corresponds to the magnetization,
but the distribution functions, and accordingly, the Binder
cumulants, are quite different.

We concentrate on the dependency of the critical Binder cumulants,
$U^{\ast}_{\rm M}$ and $U^{\ast}_{\rm l_\infty}$, on the boundary
conditions and the aspect ratio and compare our results with
previous work when available~\cite{KB1993,Selke06,BD85}. We also
illustrate and compare the corresponding probability density
functions (pdfs) observing their evolution as a function of the
aspect ratio. For PBC and $r=4$, a pronounced double-peak
structure is observed in the pdf of LFKC, and we give for this a
geometrical explanation, involving the probability that the LFKC
percolates along both (short and long) directions of the lattice
simultaneously. Finally, we discuss the critical-exponent
equivalence~\cite{Hu84,Meo} and the scaling properties of $Q=|M|$
and $Q=l_\infty$, for both PBC and FBC. In particular, we estimate
the amplitudes $A_{\rm Q}$ of the power law $\langle Q
\rangle=A_{\rm Q}L^{-\beta/\nu}$ for all values of the aspect
ratio considered. It is shown that these amplitudes follow a power
law with $r$, and the corresponding exponents are determined. This
analysis is related to the interesting superscaling concepts
reported by Watanabe \emph{et al.}~\cite{Wa04} in their study of
percolation on rectangular domains, as will be further discussed
below.

\begin{table}
\caption{\label{tab:1} Critical Binder cumulants of the order
parameter and the LFKC for PBC and FBC and several values of the
aspect ratio $r$. The numbers in parentheses denote errors.}
\begin{tabular}{lcccc}
\hline \hline
BC & $r$ & $U^{\ast}_{\rm M}\footnotemark[1]$ & $U^{\ast}_{\rm M}$ & $U^{\ast}_{\rm l_\infty}$ \\
\hline
PBC  & 1    &  $0.61069\cdots$~\cite{KB1993}  & 0.61067(24) & 0.6167(2)  \\
PBC  & 4    &  $0.48723\cdots$~\cite{KB1993}  & 0.48697(36) & 0.4995(5)  \\
PBC  & 9    &  $0.27054\cdots$~\cite{KB1993}  & 0.2713(10) & 0.3880(8)  \\
PBC  & 16    &                                 & 0.1539(6) & 0.4317(10)  \\
PBC  & 25    &                                 & 0.0984(10) & 0.4760(6)  \\
PBC  & 36    &                                 & 0.0685(10) & 0.5044(10)  \\
PBC  & 50    &  $0.04920\cdots$~\cite{KB1993}  & 0.0493(5) & 0.5258(15)  \\
PBC  & 64    &                                 & 0.0375(12) & 0.5385(8)  \\
PBC  & 100    &  $0.02454\cdots$~\cite{KB1993}  & 0.0242(20) & 0.5585(10)  \\
\hline
FBC  & 1    &  $0.396(2)$~\cite{Selke06}   & 0.3969(6) & 0.4370(10)  \\
FBC  & 4    &                              & 0.2365(15) & 0.3898(10)  \\
FBC  & 9    &                            & 0.1188(12) & 0.4394(80)  \\
FBC  & 16    &                            & 0.0680(5) & 0.4860(40)  \\
FBC  & 25    &                            & 0.0452(14) & 0.5160(20)  \\
FBC  & 36    &                            & 0.0330(6) & 0.5365(5)  \\
FBC  & 50    &                            & 0.0214(10) & 0.5533(12)  \\
FBC  & 64    &                            & 0.0178(15) & 0.5631(15)  \\
FBC  & 100    &                            & 0.0123(10) & 0.5790(10)  \\
\hline \hline
\end{tabular}
\footnotetext[1]{Best estimates from the literature.}
\end{table}

The square Ising systems under study were simulated only at the
exact critical temperature $k_{B}T_{c}/J=2.2691853\cdots$. In our
numerical approach, we define a Swendsen-Wang Monte Carlo step to
consist of $10-20$ (depending on $L^{\ast}$) Swendsen-Wang moves,
in which all Fortuin-Kasteleyn clusters attempt to flip with
probability $1/2$~\cite{Swendsen,Newman99}.
\begin{figure}[htbp]
\includegraphics*[width=8 cm]{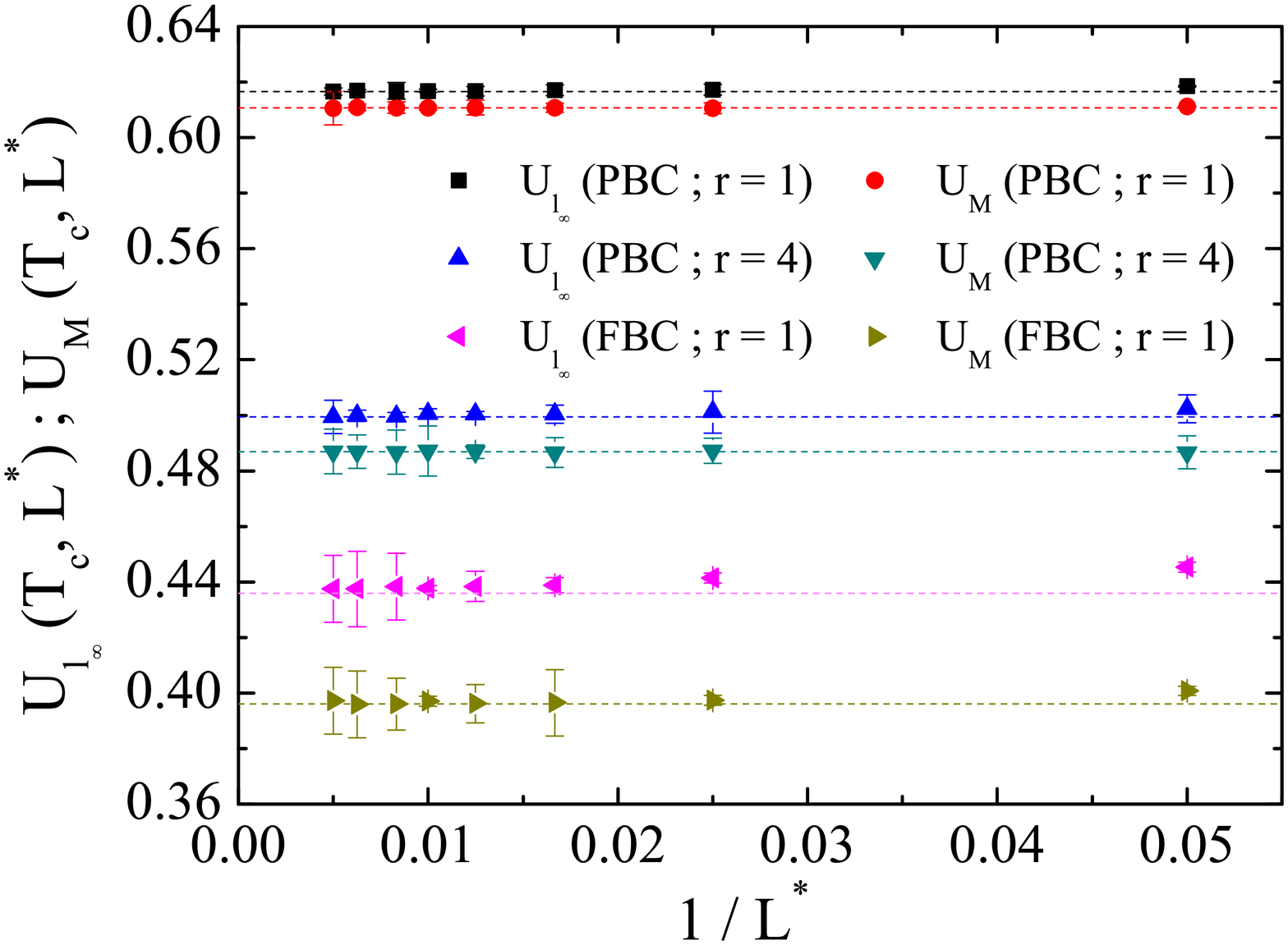}
\caption{\label{fig:1} (color online) Illustration of the
finite-size behavior of critical Binder cumulants for the
magnetization $U^{\ast}_{\rm M}$ and the LFKC $U^{\ast}_{\rm
l_\infty}$. Cases of PBC with $r=1$ and $r=4$ and FBC with $r=1$
are illustrated. The dashed lines are the extrapolated limits.}
\end{figure}
A number of Swendsen-Wang Monte Carlo steps, denoted as $n_{\rm
eq}$, is used for equilibration and a large number of such steps,
denoted as $n_{\rm rec}$, is used for the recording of the data
related to the Fortuin-Kasteleyn cluster decomposition. Typical
values of the parameters $n_{\rm eq}$ and $n_{\rm rec}$, used in
our simulations, are $n_{\rm eq}=1600$ and $n_{\rm rec}=64000$ for
$L^{\ast}=20$, whereas $n_{\rm eq}=7200$ and $n_{\rm rec}=115200$
for $L^{\ast}=120$. In each case, we used $10$ independent runs,
restarted from new random spin configurations. The statistical
errors, of the corresponding data were set equal to $3$ standard
deviations of the $10$ independent runs. In order to achieve good
accuracy in the estimation of the above mentioned amplitudes, via
an extrapolation finite-size scaling scheme, the above described
simulations were carried out for all values of aspect ratio $r$.
In all cases, approximately the range $L^{\ast}=20-120$ was
covered by $8-10$ different widths $L$, and for $r=1$ and $r=4$,
we also simulated systems with linear sizes $L=160$ and $200$.

\section{Numerical Results and Discussion}
\label{sec:3}

All our estimates for the critical Binder cumulants are given in
Table~\ref{tab:1}, together with the existing ones from the
corresponding literature. In Fig.~\ref{fig:1} we illustrate the
finite-size behavior of the critical cumulants for a selected set
of the cases (as indicated on the panel), and the rather smooth
linear extrapolation, which provides us with the limiting values
of the critical Binder cumulants. The values listed in
Table~\ref{tab:1} were obtained by applying the expected leading
correction term $aL^{-1.75}$~\cite{KB1993}.
\begin{figure}[htbp]
\includegraphics*[width=8 cm]{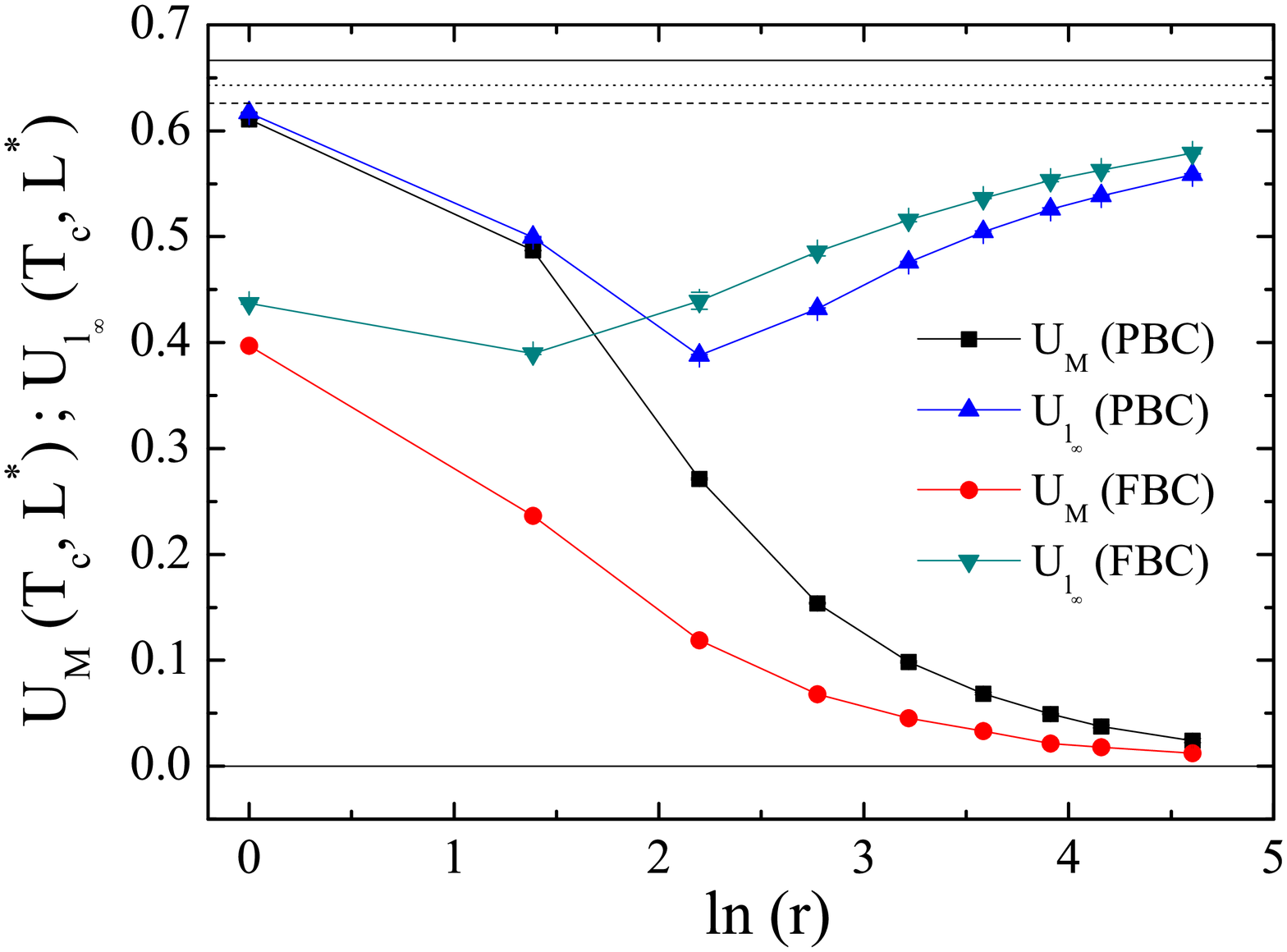}
\caption{\label{fig:2} (color online) Dependence of critical
Binder cumulants $U^{\ast}_{\rm M}$ and $U^{\ast}_{\rm l_\infty}$
on the logarithm of the aspect ratio $\ln{(r)}$ for PBC and FBC.}
\end{figure}
Note that, in almost all cases, these values and the ones obtained
by a linear extrapolation agree within error bars. From the Table
one can observe a very good agreement with previous estimates
regarding the magnetization's critical Binder cumulants and the
estimates of the present study. The critical Binder cumulants of
the LFKC ($U^{\ast}_{\rm l_\infty}$) are found, in all cases, to
be upper bounds to the values of the critical Binder cumulants of
the order parameter. The smallest difference, between the two
cumulants corresponds to the case of PBC with $r=1$ and this
difference is enhanced as the order-parameter cumulant deviates
from the value $2/3$, approaching the limiting (Gaussian) value
$0$, as $r\rightarrow \infty$.

This strong splitting behavior is presented in Fig.~\ref{fig:2},
which gives a full illustration of the dependence of critical
Binder cumulants for magnetization and the LFKC on the aspect
ratio for both cases of boundary conditions considered. Several
interesting conclusions can be drawn from this figure. First, as
should be expected, and shown by Kamieniarz and
Bl\"{o}te~\cite{KB1993} for PBC, the limiting magnetization
cumulants $U^{\ast}_{\rm M}$ agree with the Gaussian value $0$,
describing linear systems, as $r\rightarrow \infty$ for both PBC
and FBC. For large $r$, $U^{\ast}_{\rm M}(r)$ becomes linear in
$r^{-1}$, and as pointed out by Kamieniarz and
Bl\"{o}te~\cite{KB1993}, the product $A_{\rm U}(r)=rU_{\rm M}(r)$
approach exponentially fast the universal amplitude $A_{\rm
U}=\lim_{r\rightarrow \infty}{[U^{\ast}_{\rm M}(r)r]}$. The
estimates for this universal amplitude of the transfer-matrix
technique in Ref.~\cite{KB1993}, agree to $\sim 5$ significant
figures with the value $A_{\rm U}=2.46044(2)$, obtained from
conformal invariance~\cite{BD85}. The statistical Monte Carlo
errors permit here a, moderately accurate, estimate of the order
of $A_{\rm U}=2.466(7)$, as can be seen by a linear fit of the
$r=16-100$ data of Table~\ref{tab:1}. For FBC, $U^{\ast}_{\rm
M}(r)$ becomes also linear in $r^{-1}$, and the corresponding fit,
in the range $r=16-100$, provides the estimate $A_{\rm
U}=1.055(26)$.

The cumulant $U^{\ast}_{\rm l_\infty}$ of the LFKC shows a
non-monotonic behavior approaching finally, as $r\rightarrow
\infty$, a non-trivial value different to $2/3$ (describing the
ordered phase). To estimate the limiting behavior, we assume the
law $U^{\ast}_{\rm l_\infty}(r)=U^{\ast}_{\rm
l_\infty}(\infty)-Br^{-x}$. Fitting the $r=16-100$, or the
$r=25-100$ data of Table~\ref{tab:1} we find quite stable
estimates. For the $r=25-100$ range, this estimation scheme gives
$U^{\ast}_{\rm l_\infty}(r)=0.626(4)-0.96(4)r^{-0.57(2)}$ for PBC,
and $U^{\ast}_{\rm l_\infty}(r)=0.643(10)-0.63(7)r^{-0.50(6)}$ for
FBC.
\begin{figure}[htbp]
\includegraphics*[width=8 cm]{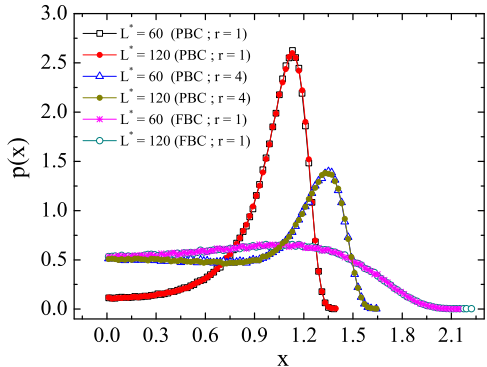}
\caption{\label{fig:3} (color online) Scaled pdfs of the
magnetization, $x=|M|/\sqrt{\langle M^2 \rangle}$, for (i) PBC
with $r=1$ and $r=4$ and (ii) FBC with $r=1$. Sizes $L^{\ast}=60$
and $L^{\ast}=120$ are illustrated.}
\end{figure}
Although even for $r=100$, the values of the cumulants deviate
significantly from their limiting values, the exponents $x$ agree
with the value $0.5$ within error bars and appear to be
independent of the boundary conditions. Thus, the exponents
controlling the limiting $r$ behavior are different for
$U^{\ast}_{\rm M}(r)$ and $U^{\ast}_{\rm l_\infty}(r)$, but are
independent of the boundary conditions. Note that, the above
limiting values are indicated by the dashed (PBC) and dotted (FBC)
lines in the panel of Fig.~\ref{fig:2} together with the full line
corresponding to the value $2/3$. In the limit $r\rightarrow
\infty$, cumulant universality between PBC and FBC is reflected in
the exponents, and the role of the boundary conditions appears to
diminish in that limit. As noted above, for moderate values of the
aspect ratio, both $U^{\ast}_{\rm M}$ and $U^{\ast}_{\rm
l_\infty}$ have a strong dependence on the boundary conditions.
This non-monotonic behavior, and the smooth final approach in the
limit $r\rightarrow \infty$, are reflections of the evolution
features of the corresponding pdfs, which are further illustrated
bellow.

\begin{figure}[htbp]
\includegraphics*[width=8 cm]{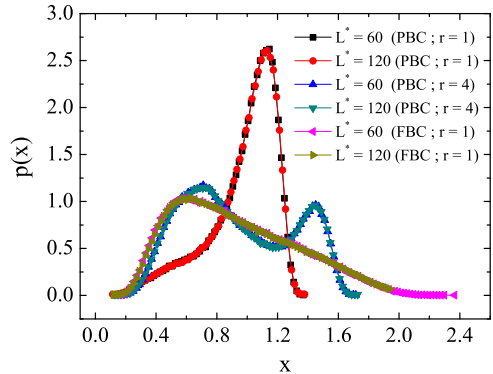}
\caption{\label{fig:4} (color online) Scaled probability pdfs of
the LFKC, $x=l_\infty/\sqrt{\langle l_\infty^2 \rangle}$,
corresponding to the cases of Fig.~\ref{fig:3}.}
\end{figure}

Figure~\ref{fig:3} illustrates the scaling of the order-parameter
pdfs, while Fig.~\ref{fig:4} illustrates the scaling of the
relevant functions of the LFKC for the cases with PBC and FBC, for
which their Binder's cumulant finite-size behavior is illustrated
in Fig.~\ref{fig:1}. The scaled distributions have been
constructed by using as scaling variables the $x=Q/\sqrt{\langle
Q^2 \rangle}$~\cite{Bind81,Bruce81,Tsypin00,mal06,Plasc07}, with
$Q=|M|$ or $Q=l_\infty$. In the scaling limit (system size going
to infinity) these functions are expected to be universal and
characterize the given universality class~\cite{Bind81,Bruce81}.
The scaled density functions are then obtained from
$p(x)dx=p_{Q}(Q)dQ$, i.e., $p(x)=p_{Q}(Q) \cdot \sqrt{\langle Q^2
\rangle}$, and also a smoothing process of the fluctuations has
been applied. The pronounced double-peak structure for PBC and
$r=4$, the left shoulders for PBC and $r=1$ observed in
Fig.~\ref{fig:4}, and also the non-monotonic behavior of the
critical cumulant of the LFKC of Fig.~\ref{fig:2}, are interesting
findings reflecting geometrical features of the present bond
correlated percolation model. A brief qualitative description of
these features is attempted below, observing the variation of
certain percolation probabilities with the aspect ratio ($r$).

\begin{figure}[htbp]
\includegraphics*[width=8 cm]{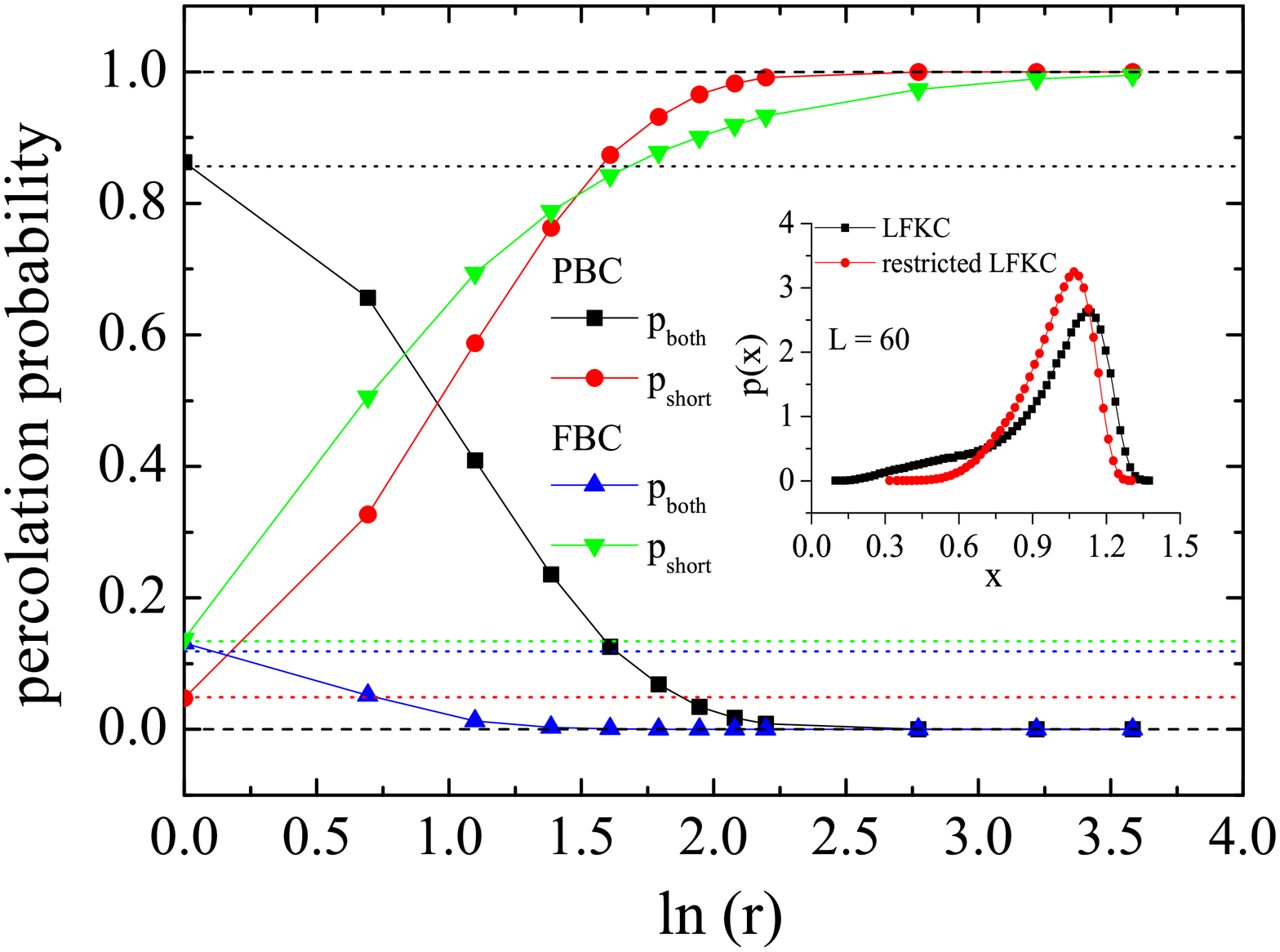}
\caption{\label{fig:5} (color online) Percolation probabilities
$p_{\rm both}$ and $p_{\rm short}$ with respect to $\ln{(r)}$.
Dashed lines indicate the $r=1$ values in the thermodynamic limit
(see discussion in the text), and the expected large-$r$ behavior.
The crossover behavior for the PBC is associated with the
appearance of the double-peak structure in Fig.~\ref{fig:4}. In
the inset we compare the $r=1$ shoulder-like behavior in
Fig.~\ref{fig:4} with a restricted pdf describing only the LFKC
that percolate simultaneously in both directions.}
\end{figure}

Throughout the years, different percolation probabilities have
appeared in the literature~\cite{Ziff92,Hu98,Cardy92}, using
different terms such as spanning probability~\cite{Ziff92},
crossing probability~\cite{Cardy92} or existence
probability~\cite{Hu98,Wa04}. It is well known that above the
percolation threshold there exists an infinite cluster with
probability one~\cite{Stauffer}, while exactly at the percolation
threshold the ``crossing probabilities'' need not be one and their
study is an important topic with many and famous contributions,
such as Cardy's exact result~\cite{Cardy92} on the square lattice
with FBC. These critical probabilities depend on the boundary
conditions and the aspect ratio~\cite{Hovi96,Wa04}. For the square
systems considered here, with $L$ rows and $L_1=rL$ columns, we
define $p_{\rm short}$ to be the probability that the LFKC
percolates only in the short direction, visiting every row of the
lattice (that is having at least one point in every row).
Respectively, the corresponding probability that the LFKC
percolates only in the long direction, visiting all columns will
be denoted by $p_{\rm long}$, and the probability of
simultaneously percolating in both the short and long directions
by $p_{\rm both}$. We may note here that for the present
bond-correlated LFKC percolation, at the critical point, the sum
$p_{\rm span}=p_{\rm both}+p_{\rm long}+p_{\rm short}$ (spanning
probability in some direction) will also depend on the boundary
conditions and the aspect ratio and need not be one. The behavior
of $p_{\rm both}$ and $p_{\rm short}$ as a function of $r$ is
illustrated in Fig.~\ref{fig:5} for both PBC and FBC. Clearly, a
strong variation with respect to the aspect ratio $r$ is observed.
To construct Fig.~\ref{fig:5} we have used systems with
approximately $3600$ lattice sites. For instance for $r=2$,
lattices of $42\times 84$ with $3528$ lattice points were used,
while for $r=36$, lattices of $10\times 360$ with $3600$ lattice
points.
Additionally, for $r=1$, we carried out a brief finite-size
scaling analysis using data in the range $L=30-100$. The resulting
limiting values are indicated with the dashed lines in
Fig.~\ref{fig:5}, and demonstrate that the finite-size behavior in
the main panel, using systems with $3600$ lattice sites, is
already a genuine representation of the behavior in the
thermodynamic limit. As can be seen, for moderate values of $r$
and PBC, the LFKC percolates with significant probabilities in
both directions of the square lattice, giving rise to the
double-peak structure. Then, as $r$ grows, the probability for
percolation along the width of the rectangular lattices increases
substantially, whereas, it declines along the length direction,
leading to the evaporation of the right peak. The crossing of the
probabilities for PBC, in the main panel of Fig.~\ref{fig:5},
gives a clear explanation of the observed double-peak structure,
while the absence of such a structure in the case of FBC, is due
the very early and large separation of the corresponding
probabilities. The presence of the left shoulders for PBC and
$r=1$, in the pdf of the LFKC shown in Fig.~\ref{fig:4}, may also
be explained by the existence of a non-vanishing contribution of
LFKC percolating only in the short direction of the lattice. This
is illustrated in the inset of Fig.~\ref{fig:5}, where we compare
this shoulder-like behavior to a restricted pdf describing only
the LFKC that percolates simultaneously in both directions.

\begin{figure}[htbp]
\includegraphics*[width=8 cm]{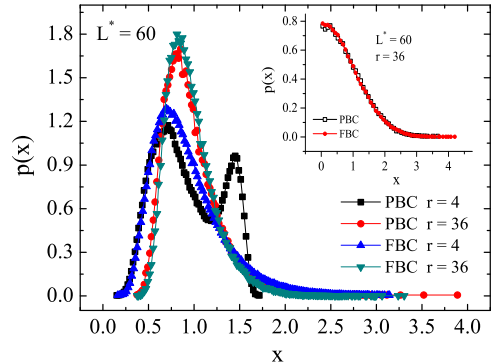}
\caption{\label{fig:6} (color online) Main panel: Scaled pdfs of
the LFKC, $x=l_\infty/\sqrt{\langle l_\infty^2 \rangle}$, for PBC
and FBC. Illustration of striking differences for $r=4$ and
approach to the same universal pdf as $r$ increases ($r=36$).
Inset: Illustration of the similarity of the magnetization's pdfs,
$x=|M|/\sqrt{\langle M^2 \rangle}$, for large $r$ ($r=36$).}
\end{figure}

Subsequently, Fig.~\ref{fig:6} clarifies in the main panel the
diversity in the shape of the pdfs of the LFKC for the cases with
PBC and FBC for moderate aspect-ratio values and the similarity in
the shape for larger values ($r=36$). The inset of the same figure
points out the similarity in the shape of the magnetization pdfs
for $r=36$.

\begin{figure}[htbp]
\includegraphics*[width=8 cm]{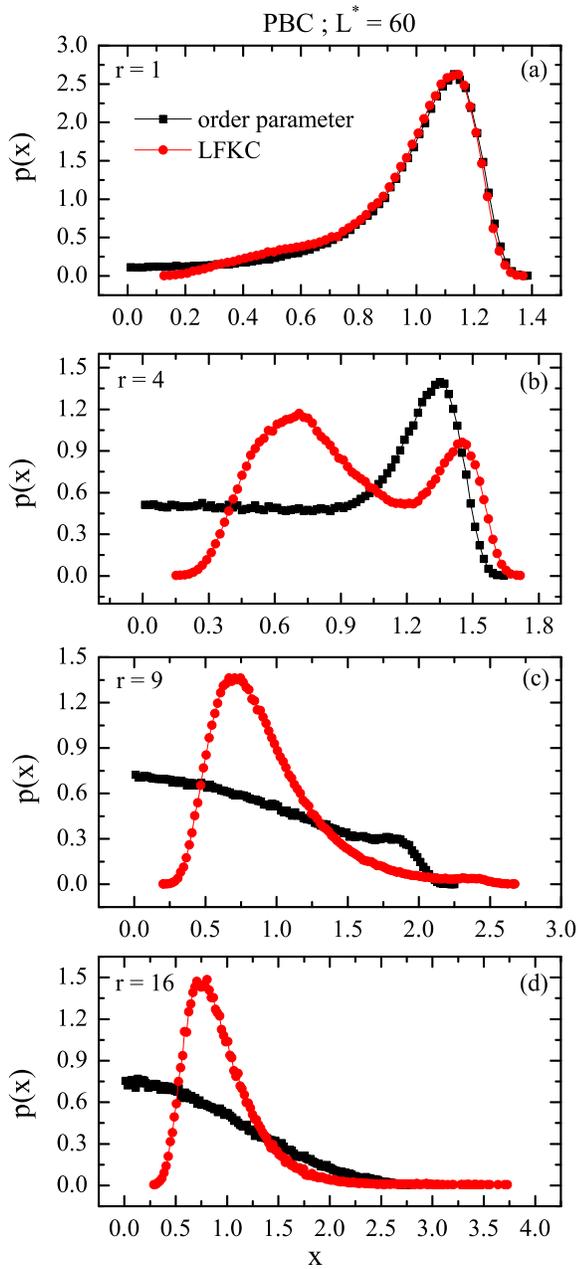}
\caption{\label{fig:7} (color online) Scaled pdfs of the
magnetization, $x=|M|/\sqrt{\langle M^2 \rangle}$, and the LFKC,
$x=l_\infty/\sqrt{\langle l_\infty^2 \rangle}$, for PBC as we vary
the aspect ratio in the window $r=1 - 16$. This figure elucidates
the similarity among the two functions for $r=1$, but also
highlights their striking differences with increasing $r$.}
\end{figure}

Differences in the shapes of the universal pdfs of the
magnetization and the LFKC should be expected from the theoretical
arguments of Hu~\cite{Hu84} and the Monte Carlo study of De Meo
\emph{et al.}~\cite{Meo}.
As shown numerically in Ref.~\cite{Meo}, below $T_c$ the
magnetization susceptibility differs from the corresponding
percolation susceptibility. Thus, the above illustrations reveal
these differences at $T_{c}$, but also show a variability in the
behavior for moderate values of the aspect ratio. The
illustrations in Fig.~\ref{fig:7} give a sketch of the evolution
of both pdfs, as we increase the aspect ratio from $r=1$ to $r=16$
in the case of PBC. From the first panel (a) of Fig.~\ref{fig:7}
for the case $r=1$, we observe a small but noticeable difference
in the left-tails, which is however enough to produce the small
difference of the cumulant values in Table~\ref{tab:1}. The
double-peak structure of the pdf of the LFKC for $r=4$, in panel
(b) of Fig.~\ref{fig:7} is associated to the pronounced left tail
in the pdfs of the magnetization. For larger values of the aspect
ratio [see panels (c) and (d) in Fig.~\ref{fig:7}] the pdfs of the
LFKC tend to the shape illustrated in the main panel of
Fig.~\ref{fig:6} for $r=36$, and the corresponding pdfs of the
magnetization to the one illustrated in the inset of
Fig.~\ref{fig:6} for $r=36$. Now, according to
Refs.~\cite{Hu84,Meo}, the behavior of $\langle Q \rangle$, with
$Q=|M|$ or $Q=l_\infty$, should be expected to be described by the
same power law of the form $\langle Q \rangle=A L^{-\beta/\nu}$,
with the critical exponent having the 2d Ising value
$\beta/\nu=0.125$. Thus, as we vary the aspect ratio, the
corresponding leading amplitudes, $A_{\rm |M|}$ and $A_{\rm
l_\infty}$, decrease and describe the asymptotic shifts to smaller
mean values. These shifts are considerable and are responsible for
the development of a Gaussian-like shape in the small $M$ behavior
of the pdfs of the magnetization, as also illustrated in the inset
of Fig.~\ref{fig:6} for $r=36$.

From the shape of the pdfs of the LFKC in Fig.~\ref{fig:4}, and
also from the comparative plot of Fig.~\ref{fig:7}, we observe
that larger fluctuations (widths) of the LFKC enhances, in all
cases, the left tail of the magnetization's pdf, since smaller
Fortuin-Kasteleyn clusters favor the mixing of positive and
negative spin clusters. However, for large values of the aspect
ratio, the pronounced $M=0$ behavior of the universal pdfs is
mainly due to the shifts of the density functions of the LFKC,
discussed above. Comparing their shapes, Figs.~\ref{fig:3} and
\ref{fig:4}, we can appreciate this evolution for the case of FBC
with $r=1$ and also for the case with PBC and $r=4$. The large
fluctuations of the LFKC and their shifts to smaller mean values
induce similar behavior on the statistically significant part of
the Fortuin-Kasteleyn clusters, which contribute and enhance the
small $M$ behavior of the order-parameter pdf.

\begin{figure}[htbp]
\includegraphics*[width=8 cm]{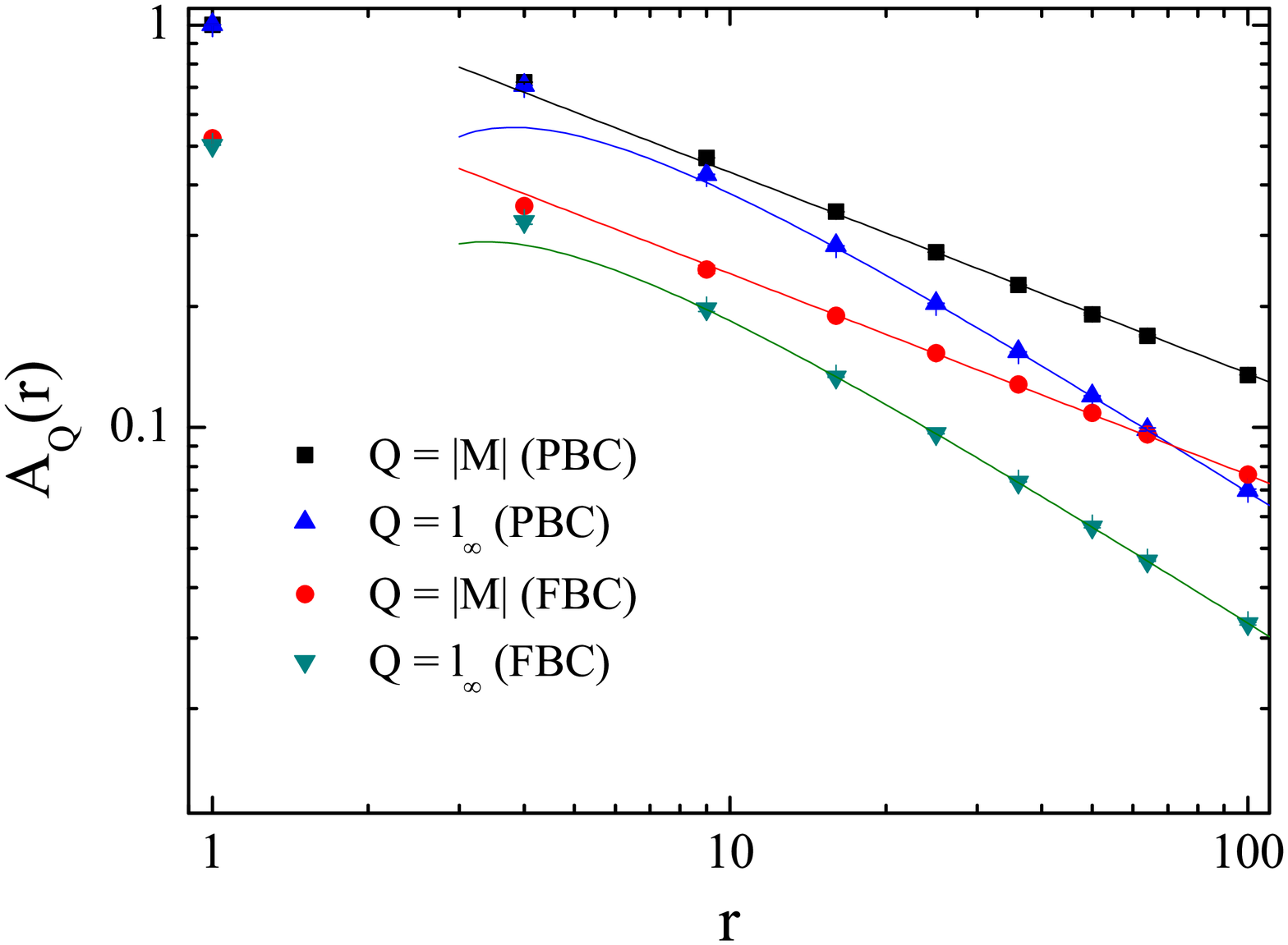}
\caption{\label{fig:8} (color online) Amplitude dependence on $r$
and illustration of the power law $A_{\rm Q}(r)=ar^{-z}$ on a
double logarithmic scale. The lines through the estimates in the
range $r \geq 3$ represent only the asymptotic behaviors, as also
discussed in the main text.}
\end{figure}

Let us now consider the scaling properties of $|M|$ and
$l_\infty$, and the critical-exponent equivalence~\cite{Hu84,Meo}.
In order to observe quantitatively the scaling properties of the
above distributions, we attempted to apply to our data a simple
power law, $\langle Q \rangle=A_{\rm Q} L^{-\beta/\nu}$, and
furthermore other expansions, including algebraic powers, but also
logarithmic terms, $\langle Q \rangle=A_{\rm Q}
L^{-\beta/\nu}(1+B\ln{(L)}/L+C/L+\cdots)$. For PBC and almost all
values of $r$, the simple power law produced stable estimates with
effective exponents converging to the 2d Ising value
$\beta/\nu=0.125$. Furthermore, by fixing the exponent to the
above expected value, and varying the width range $L=(L_{\rm
min}-L_{\rm max})$, we obtained smoothly behaving effective values
for the corresponding amplitudes $A_{\rm Q}$. Employing a linear
extrapolation in $1/L_{\rm min}$ to the above effective values, we
found very accurate estimates for the amplitudes in the
thermodynamic limit. The same procedure was also followed using
the simplest correction form $\langle Q \rangle=A_{\rm Q}
L^{-\beta/\nu}(1+B/L)$, giving final estimates which are the same,
within error bars, to those obtained by the simple power law. For
the case of FBC, the fitting attempts to the simple power law
produce, in general, overestimated values of the exponent, and by
fixing $\beta/\nu=0.125$ the resulting effective values of
amplitudes $A_{\rm Q}$, deviate significantly from their
asymptotic values. However, the fitting attempts to the form
$\langle Q \rangle=A_{\rm Q}L^{-\beta/\nu}(1+B/L)$, gave a smooth
behavior of effective amplitudes with small deviations of their
asymptotic values, allowing an accurate estimation of the
amplitudes.

\begin{table*}
\caption{\label{tab:2} Amplitudes of the power law $\langle Q
\rangle=A_{\rm Q} L^{-\beta/\nu}$ obtained by the schemes detailed
in the text using the data of magnetization and the LFKC.}
\begin{tabular}{ccccc}
\hline \hline
 & \;\;\;\;\;\;\;\;\;\;\;\;\;\;\;\;\;\;\;\;\; PBC &&
\;\;\;\;\;\;\;\;\;\;\;\;\;\;\;\;\;\;\;\;\; FBC & \\
\hline
$r$ & $A_{\rm |M|}$ & $A_{\rm l_\infty}$ & $A_{\rm |M|}$ & $A_{\rm l_\infty}$ \\
\hline
$1$ & $1.008(2)$ & $1.007(2)$ & $0.5239(4)$ & $0.5031(4)$ \\
$4$ & $0.7204(1)$ & $0.7073(1)$ & $0.3550(10)$ & $0.3250(50)$ \\
$9$ & $0.4674(2)$ & $0.4251(3)$ & $0.2470(60)$ & $0.1970(30)$ \\
$16$ & $0.3436(4)$ & $0.2826(4)$ & $0.1896(2)$ & $0.1335(2)$ \\
$25$ & $0.2727(2)$ & $0.2033(2)$ & $0.1530(4)$ & $0.0965(1)$ \\
$36$ & $0.2259(2)$ & $0.1542(2)$ & $0.1278(2)$ & $0.0732(1)$ \\
$50$ & $0.1912(12)$ & $0.1198(3)$ & $0.1088(3)$ & $0.0567(4)$ \\
$64$ & $0.1689(9)$ & $0.0988(9)$ & $0.0962(4)$ & $0.0466(2)$ \\
$100$ & $0.0764(2)$ & $0.0697(6)$ & $0.0764(8)$ & $0.0325(2)$ \\
\hline \hline
\end{tabular}
\end{table*}

In particular, for the case of PBC with $r=1$, we found by
applying a simple power law, $\beta/\nu=0.1248(4)$ from the
magnetization data and $\beta/\nu=0.1249(2)$ from the $l_\infty$
data, with corresponding amplitudes $A_{\rm |M|}=1.008(2)$ and
$A_{\rm l_\infty}=1.007(2)$. However, this was an exceptionally
good case, while for FBC with $r=1$, the simple power law, when
applied in the range $L=20 - 96$, produces the results $\langle
|M| \rangle=0.581(5) L^{-0.146(2)}$ and $\langle l_\infty
\rangle=0.556(4) L^{-0.145(2)}$. Moving to larger values of
$L_{\rm min}$, upon using the range $L=60 - 96$, we found $\langle
|M| \rangle=0.559(5) L^{-0.137(2)}$ and $\langle l_\infty
\rangle=0.541(3) L^{-0.139(1)}$. Thus, even the simple power law
improves the estimation by increasing $L_{\rm min}$. However, the
fitting attempts using the correction term are now most effective,
giving in the range $L=20 - 96$, $\langle |M| \rangle=0.520(5)
L^{-0.124(2)}(1+1.07(9)/L)$ and $\langle l_\infty \rangle=0.504(4)
L^{-0.125(2)}(1+0.98(7)/L)$. The sequence of effective estimates
resulting from the scheme with the correction term and a fixed
exponent to the expected value $\beta/\nu=0.125$, converges
smoothly to $A_{\rm |M|}=0.5239(4)$ and $A_{\rm
l_\infty}=0.5031(4)$. These results indicate that the amplitudes
of $\langle |M| \rangle$ and $\langle l_\infty \rangle$ are in
general different, and in the case of FBC the simplest correction
term $B/L$, is very effective since already the effective
estimates of the range $L=20 - 96$ are very close to their
asymptotic values.

The systematic application of the above described extrapolation
schemes verifies that for all aspect ratios $r$ the amplitude
$A_{\rm |M|}$ is higher from $A_{\rm l_\infty}$, and in fact their
difference grows with increasing $r$. This is an interesting topic
related to the theoretical arguments of Hu~\cite{Hu84} and to the
superscaling concepts reported by Watanabe \emph{et al.} in their
study of percolation on rectangular domains~\cite{Wa04}. Adapting
this superscaling proposal of Ref.~\cite{Wa04} to our study at
criticality, we assume that the above amplitudes follow for large
$r$ a power law. This is equivalent to the proposal $\langle Q
\rangle=ar^{-z}L^{-\beta/\nu}(1+\cdots)$ and thus for the relevant
amplitudes we associate a superscaling exponent $z$. Watanabe
\emph{et al.}~\cite{Wa04} have pointed out the interest of a study
of their superscaling concept to correlated percolation
models~\cite{Hu84,Hu84b}, such as the present model. In order to
verify these concepts, in the realm of the present study, we have
carried out an accurate estimation of the amplitudes for all $r$
considered here. Our estimates are given in Table~\ref{tab:2} and
the general behavior is illustrated in Fig.~\ref{fig:8}. The
amplitudes of the magnetization follow closely, for large enough
$r$, say $r\geq 9$, a simple power law of the form $A_{\rm
|M|}(r)=ar^{-z}$, with small corrections. Thus, for PBC we obtain
$A_{\rm |M|}(r)=1.361(9)r^{-0.50(2)}$, whereas for FBC $A_{\rm
|M|}(r)=0.762(9)r^{-0.50(2)}$. For the amplitudes $A_{\rm
l_\infty}$ of the LFKC, the deviations from the simple power law
are larger. Now, a correction term of the form $B/r$ stabilizes
the behavior of the effective estimates. Thus, in the case of PBC
we find $A_{\rm
l_\infty}(r)=3.01(10)r^{-0.816(14)}(1-1.71(20)/r)$, while $A_{\rm
l_\infty}(r)=1.42(5)r^{-0.816(14)}(1-1.52(22)/r)$ for the case of
FBC. The above asymptotic behaviors have been illustrated in
Fig.~\ref{fig:8} by drawing the corresponding lines through the
estimates in the range $r \geq 3$. As expected, see also
Ref.~\cite{Wa04}, the power laws apply only for systems with large
$r$.

\section{Conclusions}
\label{sec:4}

To summarize our conclusions, the amplitudes of $\langle |M|
\rangle$ and $\langle l_\infty \rangle$ are, in general, different
and depend on the boundary conditions. Their dependence on the
aspect ratio $r$ can be meaningfully described by the superscaling
concepts of Ref.~\cite{Wa04}, and by estimating the corresponding
exponents. These superscaling exponents ($z$) are certainly
different for $\langle |M| \rangle$ and $\langle l_\infty
\rangle$, but are independent of the boundary conditions. This
universality with respect to the boundary conditions appears to be
also valid, as we have showed, for the approach of the cumulants
to their limiting values for large $r$. Our illustrations of the
distribution functions allow for a better understanding of the
different behaviors of the Binder cumulants and provide an
interpretation showing the dominance of the fluctuations of the
LFKC and the importance of their shifts for the corresponding
order-parameter's universal distribution functions. Larger
fluctuations of the LFKC and their shifts to smaller values induce
similar behaviors on the statistically significant part of the
Fortuin-Kasteleyn clusters, enhancing the small order-parameter
behavior, which is mainly responsible for large deviations of the
critical cumulant from the value $2/3$ of the ordered phase. A
straightforward future challenge emerging from the current work
would be the test of the above findings for different lattice
geometries and higher dimensions.

The main issue of this work was to explain, by looking at the
geometrical sensitivity of the LFKC upon varying the boundary
conditions and the aspect ratio, the interesting behavior of
critical Binder cumulants of the order parameter. As shown, these
features are reflected in the distribution functions of the LFKC
and it should be underlined at this point that one aspect of the
fundamental achievement in the theory of equilibrium critical
phenomena, i.e., the confirmation of universality and the
calculation of critical exponents, has been obtained via the pdfs
of the main thermodynamic variables of the system at criticality.
The use of the universal character of the order-parameter pdf in
describing critical properties of models in statistical mechanics
has been shown to be quite
valuable~\cite{Bind81,Bruce81,Tsypin00,mal06,Plasc07,Plasc13}, has
been extended to the study of pure and disordered magnetic systems
and is of current interest (for a recent review and update on the
topic see Ref.~\cite{Plasc13}).

\begin{acknowledgments}
The authors would like to thank Professor Walter Selke for a
careful reading of the manuscript and useful comments, as well as
Niklas Fricke for useful correspondence, pointing out that the
fluctuations of the order parameter are probably dominated by the
size of the largest Fortuin-Kasteleyn cluster.
\end{acknowledgments}

{}
\end{document}